\pdfoutput=1 
\documentclass[12pt,preprint]{aastex}
\usepackage{natbib}
\begin{document}

\title{Discovery of Fog at the South Pole of Titan}
\author{M.E. Brown, A.L. Smith, C.Chen}
\affil{Division of Geological and Planetary Sciences, California Institute
of Technology, Pasadena, CA 91125}
\email{mbrown@caltech.edu}
\author{M. \'Ad\'amkovics}
\affil{Department of Astronomy, University of California, Berkeley, CA 94720}

\begin{abstract}
While Saturn's moon Titan appears to support an active methane hydrological
cycle, no direct evidence for surface-atmosphere exchange has yet appeared.
It is possible that the
 identified lake-features could be filled with ethane,  
an involatile long term 
residue of atmospheric photolysis; the apparent stream and channel 
features could be ancient from a previous climate; and the tropospheric
methane clouds, while frequent, could cause
no rain to reach the surface. 
We report here the detection of fog at the south
pole of Titan during late summer using observations from the VIMS
instrument on board the Cassini spacecraft. While terrestrial fog
can form from a variety of causes, most of these processes are
inoperable on Titan. 
Fog on Titan can only be caused by 
evaporation of liquid methane; the detection of fog provides 
the first direct link between surface and atmospheric methane. 
Based on the detections presented here, liquid methane appears widespread at
the south pole of Titan in late southern summer, and the hydrolgical
cycle on Titan is current active.
\end{abstract}


\section{Introduction}
Saturn's moon Titan appears to support an active methane hydrological cycle, 
with evidence for clouds \citep{griffith1998, brown-titan-2002,roe-apj-2002}, 
polar lakes \citep{stofan2007}, surface changes \citep{turtle2009}, 
and liquid carved channels \citep{radar2008}. Circulation models suggest that 
liquid methane could be predominantly confined to high latitudes 
and that methane should be seasonally transported from summer 
pole to summer pole \citep{mitchell-jgr-2008}. 
Yet little concrete evidence for the 
surface-atmosphere interactions required for this cycle has been seen. 
The clouds may produce no rain\citep{emily-southpole-2006}, the large polar lakes could be filled 
with non-evaporating ethane, rather than methane \citep{brown-ethane-2008}, 
the cause of surface 
albedo changes is totally unknown \citep{turtle2009}, 
and the carved channels could be a 
record of an ancient climate \citep{griffith2008}. 
One signature of active evaporation would be fog on Titan.
In Titan's 
atmosphere, fog can only occur when near-surface air saturates through 
direct contact with liquid methane. The detection of fog would reveal the 
presence of actively evaporating surface liquid methane 
and would point to a vigorous surface-atmosphere 
exchange in a currently active hydrological cycle.

\section{Observations}
To search for fog on Titan we examined all data publically available 
through the PDS database from the VIMS \citep{brown-vims-2004} 
instrument on the Cassini 
spacecraft. VIMS is a hyperspectral imager, obtaining near-simultaneous 
images in up to 256 channels between 1 and 5 $\mu$m. This capability, 
coupled with several strong methane absorption features in Titan's 
atmosphere throughout this spectral region, allows us sum multiple wavelength
images to construct 
synthetic filters which probe to different depths in Titan's 
atmosphere. We use identical synthetic filters already developed
and demonstrated in \citet{brown-lakeeffect-2009} and Brown (2009b). 
We also added an addition synthetic filter from the sum of all channels
from 4.95 to 5.12 $\mu$m. This region of the spectrum is transparent to the
surface and is sensitive to scattering from small particles.
Fog, if present, would appear as a bright feature 
visible in the synthetic filter which probes all the way to the surface, but 
not visible in the filters which probe only to higher altitudes. 
It will appear bright in the 5 $\mu$m filter where the surface is generally
dark but small cloud particles are bright.
Such a feature would be indistinguishable from a bright permanent 
albedo mark on the surface of Titan except that fog could be highly 
variable. 

\begin{figure}
\plotone{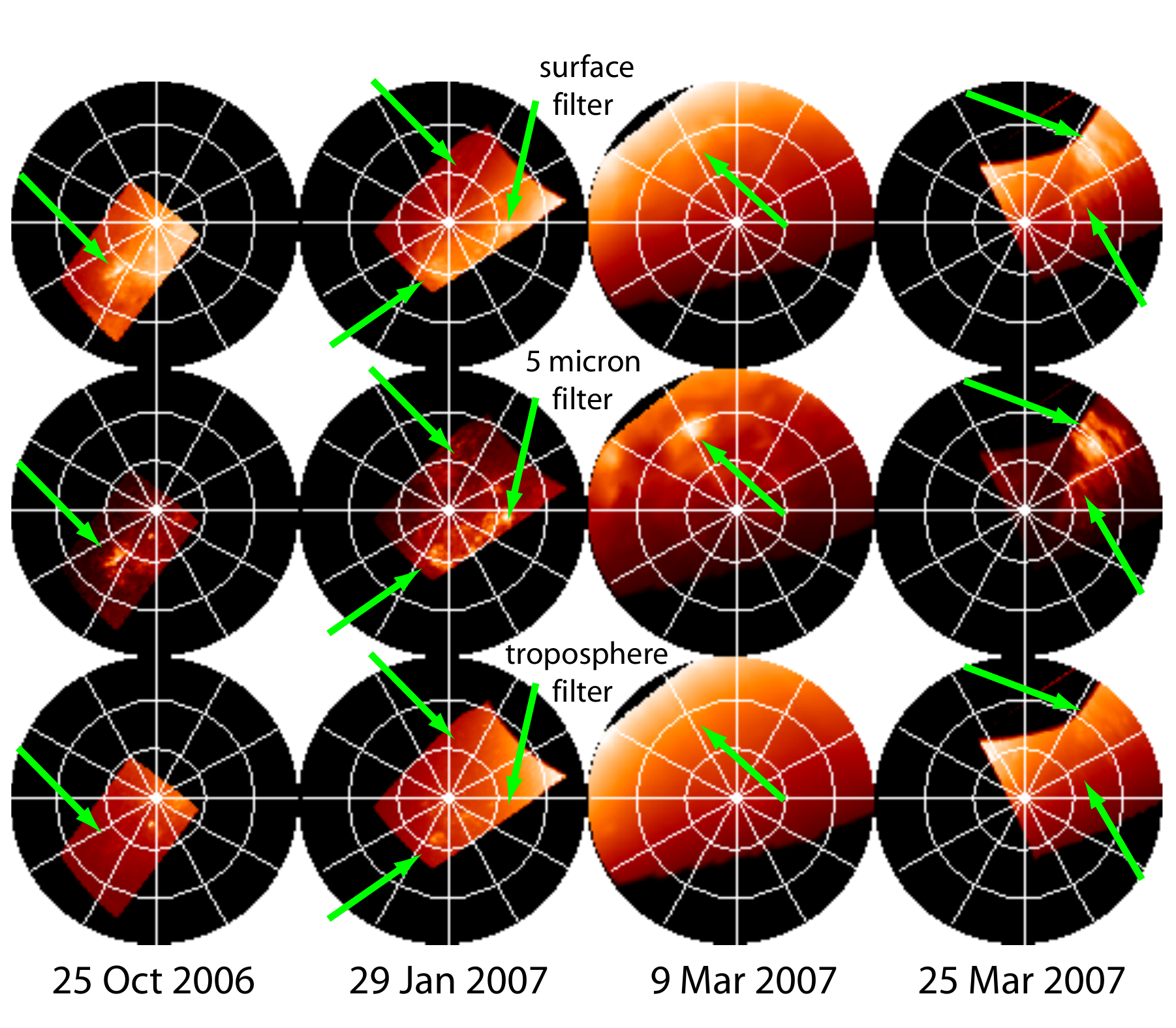}
\caption{Views of the south pole of Titan on four separate dates. Variable 
features, marked with green arrows,
 can be seen in the synthetic filter which probes to Titan's 
surface, and in the 5 $\mu$m filter, which is particularly sensitive to 
large cloud particles. These variable surface features do not appear 
in the synthetic troposphere filter, which is insensitive to scattering 
below $\sim$10km. The images are all projected to identical polar projections 
with lines of latitude between -60 and -90 shown every 10 degrees 
and with 0 degree longitude at the top.}
\end{figure}
To search for such bright variable features, we projected all 
VIMS surface and troposphere images to a common south polar 
projection and searched for the transient presence of bright surface 
features which did not appear in the tropospheric synthetic filter. 
Multiple images of a single location obtained at different 
solar and spacecraft geometries can appear subtly different; we thus 
only considered the appearance of unmistakable bright spots. Four 
such potential fog features were identified. The best examples 
are shown in Figure 1.
These images reveal the typical wispy appearance
of these features when seen at moderate 
spatial resolution. Indeed, the morphological appearance of the 25 
March 2007 image is particularly suggestive: features which appear 
in the troposphere-probing filter appear related to those which 
appear only in the surface filter. 

\begin{figure}
\plotone{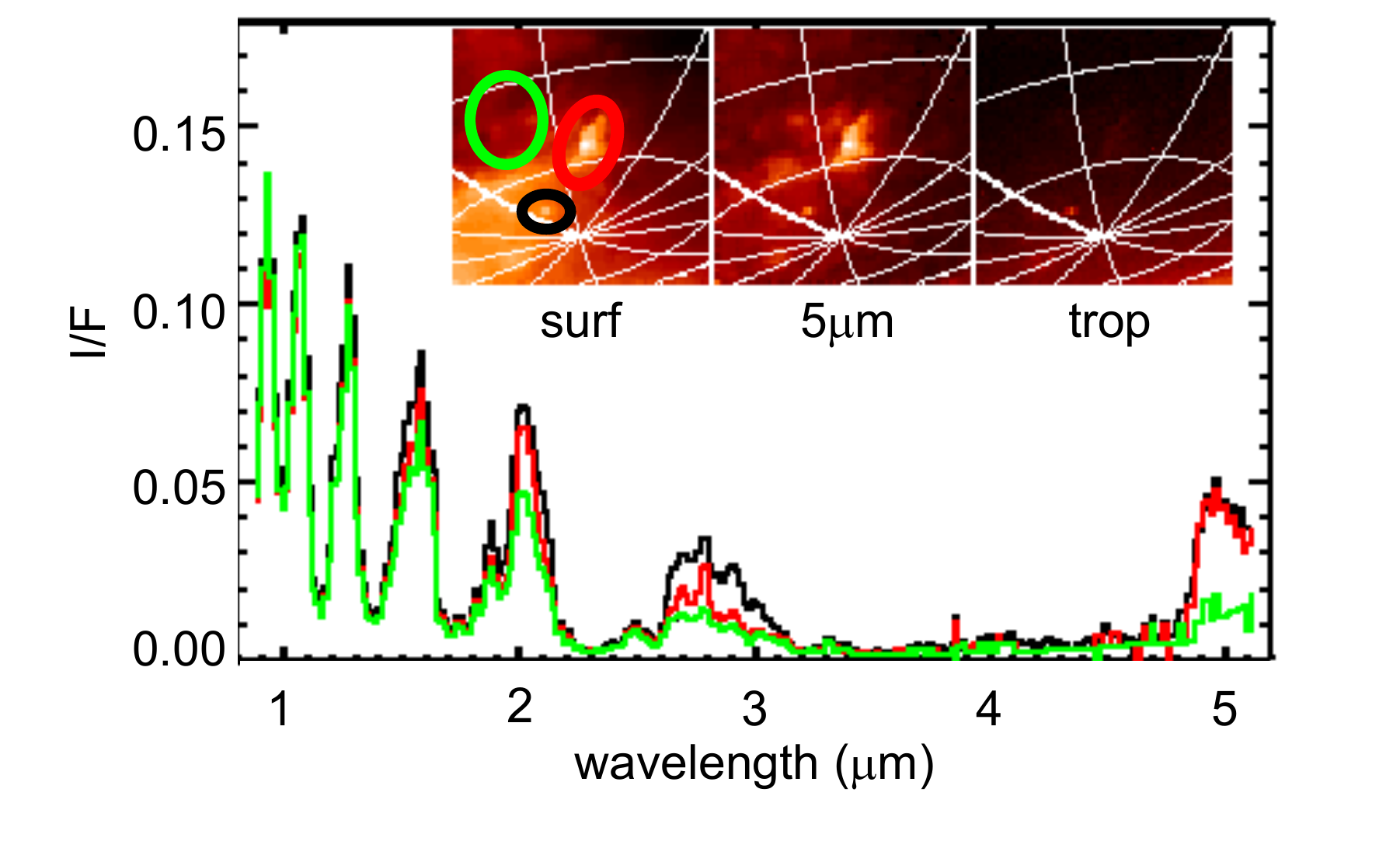}
\caption{VIMS images of 
the 25 October 2006 fog. The surface, 5 $\mu$m, and troposphere images 
use the same synthetic filters are in Figure 1. Areas of clear sky, 
fog, and tropospheric cloud are shown by green, red, and black ovals, 
respectively. The full spectra of each of these select regions are 
show in the same colors. The surface, fog, and cloud spectra clearly 
differ. Both the fog and the cloud are bright at 5 $\mu$m, while the fog 
appears intermediate between the surface and troposphere in the 
2.5-3 $\mu$m region.}
\end{figure}
To further explore the cause 
of these variable features, we examined the full 
1 - 5 $\mu$m spectra of regions of the regions ofpotential fog. Figure 2 
shows an example of comparisons between the spectra of regions that we 
identify as unvarying surface, tropospheric cloud, and the variable 
near-surface feature. 
The variable surface features appear spectrally unlike any surface unit
at the south pole.
In spectral regions that are transparent all the way to 
the surface, the variable -surface feature appears identical 
to the tropospheric cloud, including, most dramatically, the high 
reflectivity near 5 $\mu$m, when compared to the surface feature. 
Tropospheric clouds are bright at these wavelengths because they 
are composed of bright single scattering particles with sizes larger 
than the wavelength of the light \citep{barnes2005, griffith2005}.  
In spectral regions where 
light transmission to the surface is significantly attenuated, but 
where transmission to the troposphere is high 
(the 2.1 $\mu$m region, for example), tropospheric clouds appear bright
while the surface is dark. The
variable feature and the unvarying surface appear 
spectrally similar, suggesting that the variable feature originates 
from close to the surface.

\section{Analysis}
To determine the altitude of the variable 
near-surface feature, we perform full calculations of the radiative transfer 
through Titan's atmosphere using the method of  
\citet{adamkovics2007}. While accurate 
radiative transfer calcuations through the
poorly known south polar atmosphere on Titan are fraught with uncertainty,
we side-step many of these difficulties by instead performing 
simple comparisons of adjacent areas of the image with and
without fog features. Assuming that the large-scale
atmospheric scattering and opacity does not change
significantly between these regions, which are only 400 km apart, we can
accurately model the relative effect of adding fog to the spectrum.

First, the aerosol opacity profile through Titan's south polar atmosphere is
scaled until the cloud-free spectrum beyond 2.15 $\mu$m is 
reproduced. 
The cloud-free spectrum is matched below 2.15 $\mu$m by 
setting the surface reflectivity spectrum to match the observations. 
In the 2 $\mu$m window, the base surface reflectivity of 0.096 must
be attenuated 
with 5 arbitrary Gaussian features to produce the agreement 
between model and 
spectrum. This parameterization of the surface reflectivity spectrum 
is not unique, but provides a basis for comparing the spectra at adjacent 
locations.

After matching the cloud-free spectrum, a series of spectra are calculated 
with clouds of varying opacity and altitude. Clouds are modeled by 
splitting the nearest atmospheric layer and inserting a uniformly 
scattering cloud layer between them. This method maintains the aerosol 
opacity structure while including a cloud layer with distinct 
scattering properties. In the limit of very thin clouds, we confirm 
that the additional layer does not create a change in the calculated 
spectrum. Cloud particles are assumed to be large and uniformly 
scattering at all wavelengths with an albedo of 0.99 and a 
Henyey-Greenstein scattering parameter, 0.85. The best agreement 
between the models and observations occurs for a cloud layer with a 
scattering optical depth of 0.25, a scattering altitude near 0.75 km, 
and an underlying surface reflectivity of 0.10. A series of spectra 
are calculated with optically thin clouds (optical depth of $\tau$=0.25) 
in each model layer.

\begin{figure}
\plotone{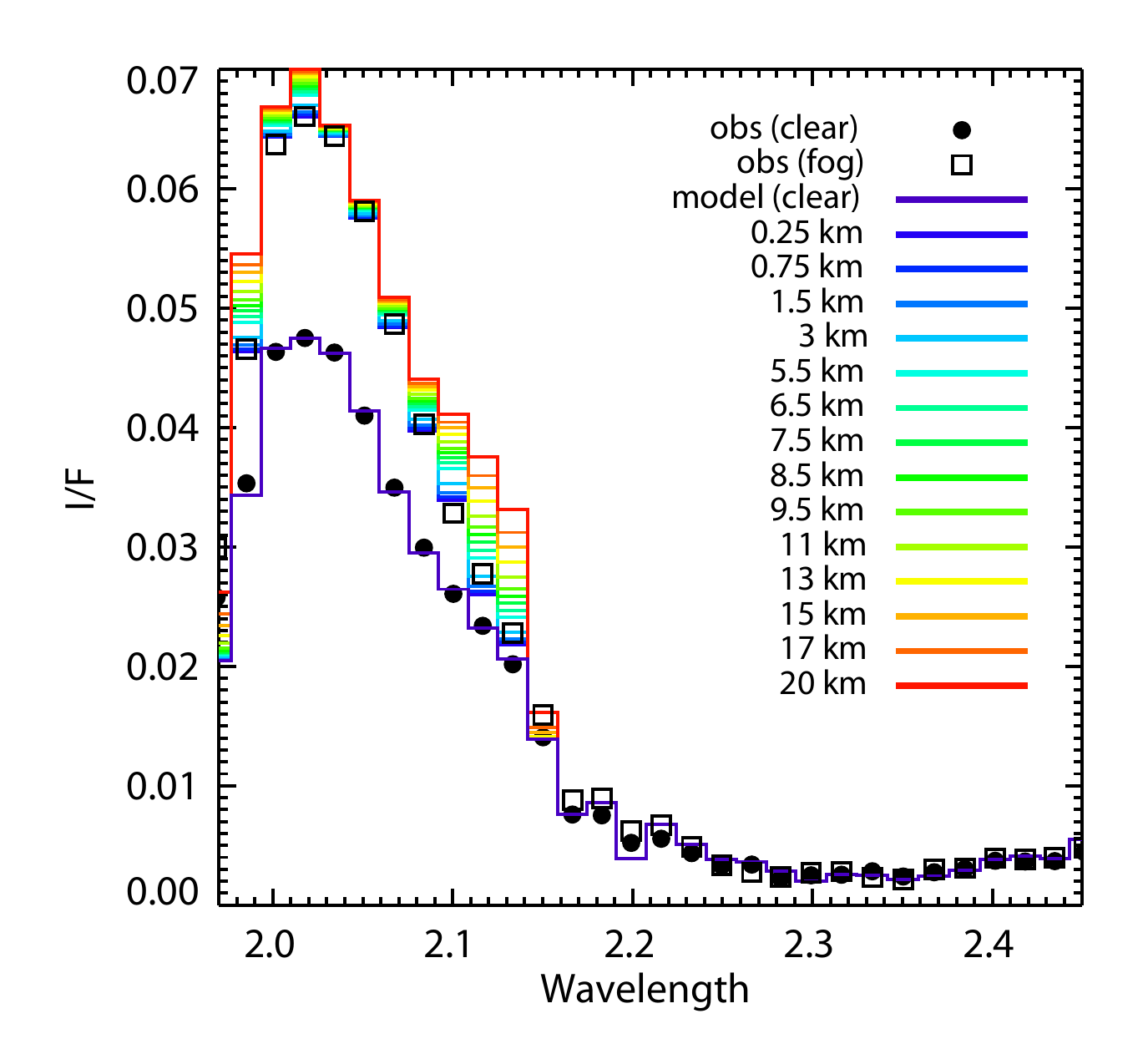}
\caption{Spectra of the surface (filled circles) 
and fog (black open squares) in the 2 $\mu$m spectral window where 
altitude is best constrained. The purple line shows the best fit 
radiative transfer model which matches the surface. Colored lines 
show fits to the fog spectra that include an increased surface 
reflectivity and a scattering cloud layer at altitudes between 
0.25 and 20 km. Cloud altitudes near 750m most closely match the 
observations, whereas models with cloud tops above 3 km altitude are 
inconsistent with the observed spectra. }
\end{figure}
As seen in Figure 3, the 
spectrum of the variable feature is best fit with a cloud with an 
altitude of 750m. Models with cloud heights of 3 km or higher differ 
significantly from the date in the 1.98 $\mu$m and the 2.09-2.14 $\mu$m regions 
which are particularly sensitive the lower troposphere. We thus conclude 
that the variable feature is indeed best described as surface fog on Titan.

\section{Discussion}
Fog forms when the vapor in ground-level air saturates and condenses. 
On the earth, this saturation can commonly occur when air radiatively 
cools overnight until it reaches the dew point. On Titan, such a 
formation mechanism is impossible. Titan's lower atmosphere has a 
radiative time constant of $\sim$100 yr \citep{hunten-titan-1984} and 
94K air that has a relative 
humidity of $\sim$50\% must be cooled $\sim$7K 
before condensation will initiate. 
Similarly, advection of typical Titan air over even the coldest locations 
on Titan provides insufficient cooling for fog to form.

Fog on Titan instead requires elevated surface humidity. If the fog 
is at ground level, the surface relative humidity must be nearly 100\%. 
Such an elevated surface humidity occurs if the surface air is in 
contact with nearly pure evaporating liquid methane.
Saturated low-level 
air on Titan is, however, unstable to moist convection for a typical 
Titan thermal profile\citep{griffith2008}. 
Fog can only persist at the surface if the 
surface air is both saturated and colder than its surroundings.

Pools of evaporating liquid methane will indeed be cooler than their 
surroundings \citep{mitri2007} and, under the right meteorological conditions, 
will 
add humidity to and drain heat from overlying air parcels.  No other 
formation explanation can naturally explain both the increased humidity 
and decreased temperature required. This formation mechanism also 
naturally explains the occasional correlation between fog and overlying 
tropospheric clouds. If surface humidity is raised without a sufficient 
decrease in temperature, saturated air parcels will convect into the 
upper troposphere \citep{griffith2000}. 
Much of the continued south polar cloudiness may 
be tied directly to surface liquid methane.

\begin{figure}
\plotone{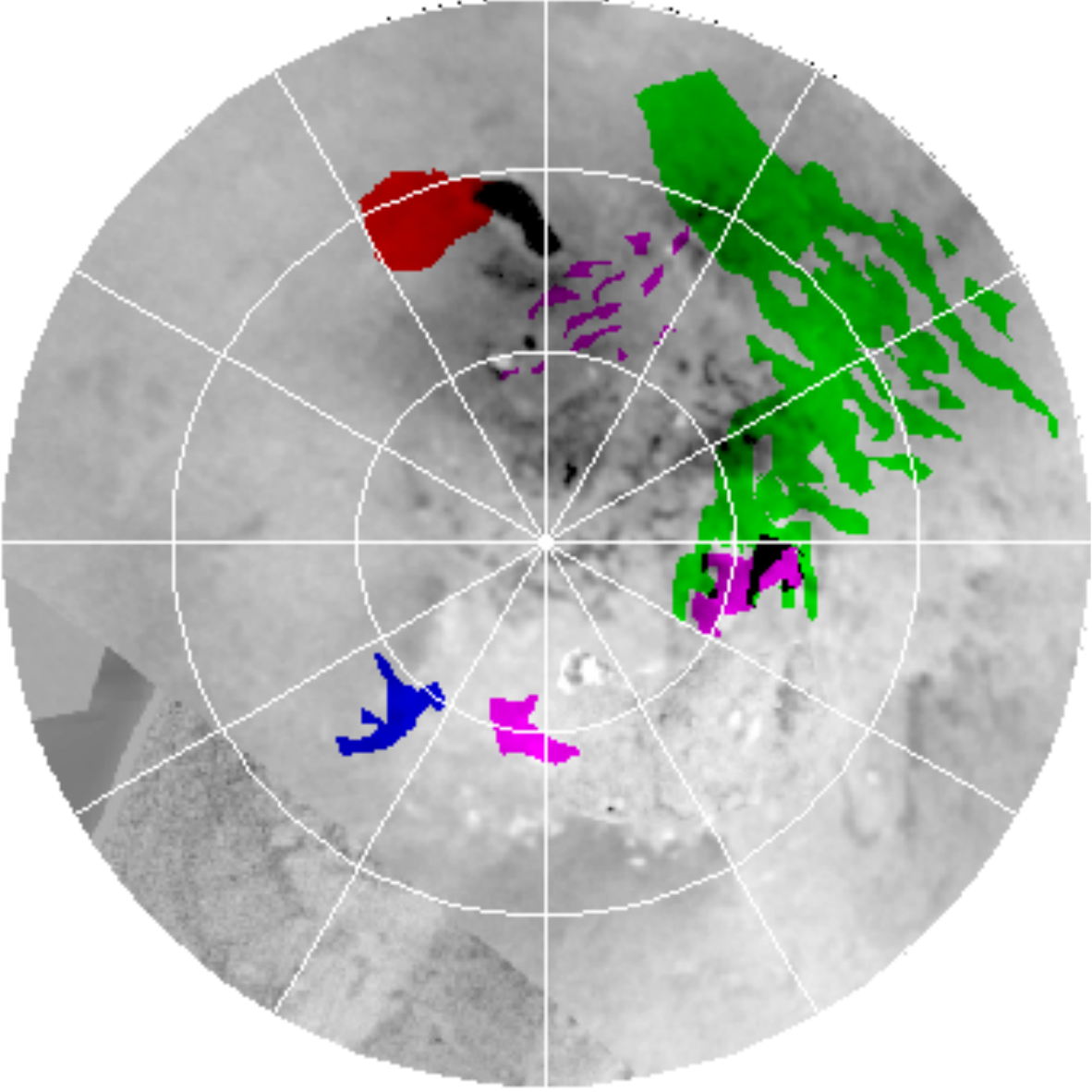}
\caption{Locations of all identified clouds figures. Blue, purple, red, 
and green are from 25 Oct 2006, 29 Jan 2007, 9 Mar 2007, and 25 Mar 2007, 
respectively. The background shows a south polar basemap of Titan as 
derived from the visible imager on the ISS instrument.}
\end{figure}

The locations of the 
identified fog features are shown in Figure 4. All identified fog 
features are southward of 65S. No correlation is seen between the 
locations of fog and the location of the one suspected large lake, 
Ontario Lacus, the locations of dark albedo features, or the location 
of the large observed albedo change \citep{turtle2009}. 
No temporal association with 
known south polar tropospheric cloud outburst appears \citep{emily-southpole-2006, emily-tropics-2009}

Fog is likely a more common occurrence than shown here; it is difficult 
to identify in the typical low resolution images obtained by VIMS, 
but seen in nearly all high resolution south polar images. Liquid 
methane and evaporation are likely even more distributed; all 
evaporation need not cause fog: special meteorological conditions 
such as low winds are also likely required. Liquid methane appears 
widespread at the south pole of Titan in the late southern summer.

\acknowledgements This research has been supported by a 
grant from the NSF Planetary Astronomy program.

\end{document}